\documentclass[prd,twocolumn,aps,psfig,nofootinbib,nobibnotes,superscriptaddress,preprintnumbers,times]{revtex4}
\usepackage[dvipdfm]{graphicx}
\usepackage{bmpsize}
\usepackage{bm,color,braket}
\usepackage{amsmath}
\graphicspath{{./fig/}{./png/}}

\newcommand{\BoldVec}[1]{\mathchoice%
  {\mbox{\boldmath $\displaystyle     #1$}}%
  {\mbox{\boldmath $\textstyle        #1$}}%
  {\mbox{\boldmath $\scriptstyle      #1$}}%
  {\mbox{\boldmath $\scriptscriptstyle#1$}}%
}
\newcommand{\EQ}{\begin{equation}}
\newcommand{\EN}{\end{equation}}
\newcommand{\EQA}{\begin{eqnarray}}
\newcommand{\ENA}{\end{eqnarray}}

\newcommand{\Eq}[1]{Eq.~(\ref{#1})}

\newcommand{\Eqss}[2]{Eqs~(\ref{#1})--(\ref{#2})}

\newcommand{\Fig}[1]{Fig.~\ref{#1}}

\newcommand{\Tab}[1]{Table~\ref{#1}}

{}
{}
{}

\newcommand{\xx}{\BoldVec{x}{}}

\newcommand{\uu}{\BoldVec{u} {}}

\newcommand{\BB}{\BoldVec{B} {}}

\newcommand{\AAA}{\BoldVec{A} {}}

\newcommand{\JJ}{\BoldVec{J} {}}

\newcommand{\ee}{\BoldVec{e} {}}

\newcommand{\hh}{\BoldVec{h} {}}

\newcommand{\kk}{\BoldVec{k} {}}

\newcommand{\nab}{\BoldVec{\nabla} {}}

\newcommand{\PPP}{{\sf P}}
\newcommand{\eee}{{\sf e}}
\newcommand{\hhh}{{\sf h}}

\newcommand{\SSSS}{\bm{\mathsf{S}}}

\newcommand{\DD}{{\rm D} {}}

\def\EEM{{\cal E}_{\rm M}}

\def\EEGW{{\cal E}_{\rm GW}}

\def\EGW{E_{\rm GW}}

\def\EM{E_{\rm M}}

\def\xiM{\xi_{\rm M}}

\def\vmu{v_{\mu}}
\def\vlam{v_{\lambda}}

\def\half{{\textstyle{1\over2}}}

\newcommand{\kHz}{\,{\rm kHz}}

\newcommand{\s}{\,{\rm s}}

\newcommand{\km}{\,{\rm km}}

\newcommand{\Mpc}{\,{\rm Mpc}}

\newcommand{\CPI}{{\text{\sc cpi}}}

\usepackage{nicefrac}
\usepackage[hidelinks]{hyperref}

\begin{document}

\title{Relic Gravitational Waves from the Chiral Plasma Instability \\ in the Standard Cosmological Model}

\date{\today}
\preprint{NORDITA-2023-034}

\author{Axel~Brandenburg}
\email{brandenb@nordita.org}
\affiliation{Nordita, KTH Royal Institute of Technology and Stockholm University, 10691 Stockholm, Sweden}
\affiliation{Department of Astronomy, AlbaNova University Center, Stockholm University, 10691 Stockholm, Sweden} 
\affiliation{School of Natural Sciences and Medicine, Ilia State University, 0194 Tbilisi, Georgia}
\affiliation{McWilliams Center for Cosmology and Department of Physics, Carnegie Mellon University, Pittsburgh, PA 15213, USA}

\author{Emma Clarke \footnote{Corresponding author; the authors are listed alphabetically.}}
\email{emmaclar@andrew.cmu.edu}
\affiliation{McWilliams Center for Cosmology and Department of Physics, Carnegie Mellon University, Pittsburgh, PA 15213, USA}

\author{Tina~Kahniashvili}
\email{tinatin@andrew.cmu.edu}
\affiliation{McWilliams Center for Cosmology and Department of Physics, Carnegie Mellon University, Pittsburgh, PA 15213, USA}
\affiliation{School of Natural Sciences and Medicine, Ilia State University, 0194 Tbilisi, Georgia}
\affiliation{Abastumani Astrophysical Observatory, Tbilisi, GE-0179, Georgia}

\author{Andrew J. Long}
\email{andrewjlong@rice.edu}
\affiliation{Department of Physics and Astronomy, Rice University, 6100 Main St., Houston, TX 77005, USA}

\author{Guotong Sun}
\email{guotongs@andrew.cmu.edu}
\affiliation{McWilliams Center for Cosmology and Department of Physics, Carnegie Mellon University, Pittsburgh, PA 15213, USA}

\begin{abstract}
In the primordial plasma, at temperatures above the scale of electroweak
symmetry breaking, the presence of chiral asymmetries is expected to
induce the development of helical hypermagnetic fields through the
phenomenon of chiral plasma instability.
It results in magnetohydrodynamic turbulence due to the high conductivity
and low viscosity and sources gravitational waves that survive in the
universe today as a stochastic polarized gravitational wave background.
In this article, we show that this scenario only relies on Standard
Model physics, and therefore the observable signatures, namely the relic
magnetic field and gravitational background, are linked to a single
parameter controlling the initial chiral asymmetry.
We estimate the magnetic field and gravitational wave spectra, and
validate these estimates with 3D numerical simulations.
\end{abstract}

\maketitle

\section{Introduction} 
The excess of matter over antimatter on cosmological scales in the universe today is well measured but its origin is not yet established. 
In studies of early universe cosmology, it is typically assumed that the matter-antimatter asymmetry arose dynamically in the first fractions of a second after the Big Bang through a process called baryogenesis \cite{Kolb:1979qa}.  
In addition to creating the baryon asymmetry, e.g., the excess of nuclei
over anti\-nuclei, baryogenesis may have created other (possibly unstable)
particle asymmetries as well; a few examples include lepton asymmetry
\cite{Fukugita:1986hr}, Higgs asymmetry \cite{Servant:2013uwa}, neutrino
asymmetry \cite{Dick:1999je,Murayama:2002je}, and right-chiral electron
asymmetry \cite{Campbell:1990fa}.
Some of these are examples of chiral asymmetries, $n_5 = n_R - n_L$, namely an excess (or deficit) of right-chiral particles and antiparticles over their left-chiral partners.  
A particular linear combination of various particle asymmetries, which we call the hypercharge-weighted chiral asymmetry, has attracted interest because of its connections with primordial magnetogenesis \cite{Joyce:1997uy} through a phenomenon known as the chiral plasma instability
\cite{Akamatsu:2013pjd}.  

The primordial magnetic field may survive in the universe today as an
intergalactic magnetic field, thereby opening a pathway to test this
scenario \cite{Neronov:2010gir,Vachaspati:2020blt}.
In addition, the primordial magnetic field and its interaction
with the turbulent plasma are expected to source gravitational
radiation; see Ref.~\cite{Deryagin:1986qq} for pioneering work
and Ref.~\cite{Brandenburg:2021aln} for numerical simulations of the
gravitational waves induced by the primordial magnetic field originating
from the chiral plasma instability.
The production of magnetic fields (possibly dark fields)
and gravitational wave radiation has also been extensively
explored in a different class of theories where the role of the
chemical potential is played by axions or axion-like particles
\cite{Anber:2009ua,Barnaby:2012xt,Domcke:2016bkh,Machado:2018nqk}.
In our work, we investigate the gravitational wave signatures of a
primordial hypercharge-weighted chiral asymmetry via the chiral plasma
instability.

Contrary to earlier numerical simulations, we study here a parameter
regime that is more realistic in various respects.
The resulting gravitational wave energy from our simulations confirm
the scaling with the sixth power of the chiral chemical potential
and the fifth power of the inverse square root of the chiral dilution
parameter, which can be combined into a single parameter, as already
found previously \cite{Brandenburg:2021aln}.

\section{Description of the model} 
We consider the primordial Standard Model plasma at temperatures $T \gtrsim 100 \ \mathrm{TeV}$ in the phase of unbroken electroweak symmetry.  
We remain agnostic as to the physics of baryogenesis, but assume that a
nonzero hypercharge-weighted chiral asymmetry is present in the plasma
initially.
We study the growth of an initially vanishingly small
hypermagnetic field via the chiral plasma instability
and calculate the resulting gravitational wave radiation.  
The hypermagnetic field generated by the chiral plasma instability is
always maximally helical and therefore also leads to the production of
maximally circularly polarized gravitational waves. 
The present work is conceptually different from that of
Refs.~\cite{Brandenburg:2023rul, Brandenburg:2023aco}, where a helical
magnetic field was present initially such that the net chirality of the 
system was balanced to zero by a fermion chirality of opposite sign.

One appealing aspect of our approach is its minimalism: we only assume Standard Model particle physics and the standard cosmological model after reheating.  
Our only free parameter is the initial hypercharge-weighted chiral asymmetry, which presumably arises from physics beyond the Standard Model.  
We work in the Lorentz-Heaviside unit system with $\hbar = c = k_B = 1$.  
We account for the cosmological expansion using an Friedmann-Lema\^itre-Robertson-Walker metric with dimensionless scale factor $a(t)$ and set $a_0 = 1$ today.  
Unless otherwise specified, all dimensionful variables are comoving; this includes 
conformal time $\mathrm{d}t = \mathrm{d}t_\mathrm{phys}/a$, 
comoving magnetic field ${\bm B} = a^2 {\bm B}_\mathrm{phys}$, 
comoving magnetic correlation length $\xiM = \xi_{\mathrm{M, phys}} / a$, 
comoving temperature $T = a T_\mathrm{phys}$, 
comoving wave number $k = a k_\mathrm{phys}$, 
comoving Hubble parameter $H = a H_\mathrm{phys}$ 
(with $H \equiv (da/dt)/a$), 
and comoving energy density of any relativistic component (including frozen-in magnetic fields, gravitational waves, etc)
$\mathcal{E} = \mathcal{E}_\mathrm{phys} a^4$.
We denote Newton's gravitational constant by $G$, the Planck mass by $M_\mathrm{Pl} = 1 / \sqrt{G} = 1.2 \times 10^{19} \, \mathrm{GeV}$,
the physical Hubble constant by $H_{\mathrm{phys},0} = 100 h_0 \, \mathrm{km}/\mathrm{sec}/\mathrm{Mpc}$, 
and the critical energy density today 
by $\mathcal{E}_\mathrm{cr} = 3H_0^2/(8\pi G)$.  
We use the subscript ``CPI'' to denote the time when the chiral plasma instability (CPI) develops.  
Assuming that the plasma's entropy density is conserved between the CPI epoch and today leads to the relation $g_{\ast S,\CPI} a_\CPI^3 T_{\mathrm{phys},\CPI}^3 = g_{\ast S,0} a_0^3 T_{\mathrm{phys},0}^3$.  
Taking $g_{\ast S,0} = 3.91$ and $T_{\mathrm{phys},0} = 0.234 \, \mathrm{meV}$ gives 
\begin{equation}\label{eq:a_CPI}
	\frac{a_\CPI}{a_0} 
	= (8 \times 10^{-19}) 
	\bigg( \frac{g_{\ast S,\CPI}}{106.75} \bigg)^{-1/3} 
	\bigg( \frac{T_{\mathrm{phys},\CPI}}{100 \, {\rm TeV}} \bigg)^{-1}
	\;.
\end{equation}
We fiducialize the effective number of relativistic degrees of freedom
during the CPI epoch to $g_{\ast S,\CPI} = 106.75$, which is the
expected value for Standard Model cosmology at temperatures above
$100 \, \mathrm{GeV}$.
We fiducialize the physical plasma temperature at the CPI epoch to
$T_{\mathrm{phys},\CPI} = 100 \, \mathrm{TeV}$.

\paragraph{\textbf{Chiral magnetic effect.}} 
The chiral plasma instability 
and chiral magnetic effect (CME) were first studied in the context of a relativistic electron-positron plasma described by quantum electrodynamics (QED). 
Although chirality is conserved at the classical level for massless electrons, chirality is broken in the quantum theory and this is expressed by the Adler-Bell-Jackiw axial anomaly \cite{Bell:1969ts,Adler:1969gk}.  
A manifestation of the anomalous chiral symmetry is the CME \cite{Vilenkin:1980fu}: in a QED plasma that possesses a chiral asymmetry, a magnetic field induces a proportional current.  
The CME corresponds to an anomalous contribution to the electric current
density ${\bm J}({\bm x},t) = \mu_5(t) {\bm B}({\bm x},t)$ where
$\mu_5 = 2 \alpha \tilde{\mu}_5 / \pi$ is proportional to the
chiral chemical potential $\tilde{\mu}_5$, 
$\alpha = e^2/4\pi \approx 1/137$ is the electromagnetic fine structure
constant, 
and ${\bm B}$ is the magnetic field.
Implications of the CME for a turbulent QED plasma have been studied extensively with a combination of analytical techniques and numerical simulations \cite{Boyarsky:2011uy,Boyarsky:2012ex,Boyarsky:2015faa,Brandenburg:2017rcb,Brandenburg:2021aln,Brandenburg:2023aco};
see also Ref.~\cite{Kamada:2022nyt} for a recent review article.  

\paragraph{\textbf{Adaptation to hypercharge.}} 
The formalism used to study the CME in QED is easily adapted to the hypercharge sector of the Standard Model for a plasma in the phase of unbroken electroweak symmetry at temperatures $T_\mathrm{phys} \gtrsim 100 \ \mathrm{GeV}$.  
The quantity of interest is the hypercharge-weighted
chiral chemical potential $\tilde{\mu}_{Y,5}(t)$, which is given by
$\tilde{\mu}_{Y,5}(t) = \sum_i \varepsilon_i g_i Y_i^2 \tilde{\mu}_i(t)$, where
the sum runs over all Standard Model particle species (indexed
by $i$), $\varepsilon_i = \pm 1$ for right/left-chiral particles
(and $0$ otherwise), $g_i$ is a multiplicity factor (counting color,
spin, etc), $Y_i$ is the hypercharge of species $i$, and $\tilde{\mu}_i$ is
the chemical potential that parameterizes the asymmetry
(excess of particles over antiparticle partners) in species $i$ via
$n_i \propto \tilde{\mu}_i T^2$; see Refs.~\cite{Kamada:2016eeb,Kamada:2016cnb}
for additional details.  

\paragraph{\textbf{Chiral plasma instability.}} 
In the presence of a chiral asymmetry, the equations of magnetohydrodynamics (MHD)
are modified due to the CME, and the new equations exhibit a tachyonic instability toward the growth of long-wavelength modes of the magnetic field, which is known as the
chiral plasma instability \cite{Akamatsu:2013pjd}.  
To illustrate the instability in the hypercharge sector of the primordial plasma, we present the evolution equation for the hyper-magnetic field assuming negligible plasma velocity:  $\dot{\bm B}_Y = \eta_Y \nabla^2 {\bm B}_Y + (2 \alpha_Y \tilde{\mu}_{Y,5} / \pi) \eta_Y {\bm \nabla} \times {\bm B}_Y$.  
Here and below, dots represent partial derivatives with respect to
conformal time, ${\bm B}_Y({\bm x},t)$ is the hypermagnetic field,
$\eta_Y = 1/\sigma_Y$ is the hypermagnetic diffusivity, $\sigma_Y$ 
is the hypercharge conductivity, and $\alpha_Y = g^{\prime 2}/4\pi
\approx 0.01$ is the hypercharge fine structure constant.
Long-wavelength modes of the hypermagnetic field with wave number $k < k_\CPI = 2 \alpha_Y |\tilde{\mu}_{Y,5}(t)|/\pi$ experience a tachyonic instability in one of the two circular polarization modes, and their amplitude increases exponentially $\propto \mathrm{exp}(t / t_\CPI)$.  
The fastest growing modes have $k = k_\CPI / 2$, and for these modes $t_\CPI = 4 / \eta_Y k_\CPI^2 = \pi^2 / \eta_Y \alpha_Y^2 |\tilde{\mu}_{Y,5}|^2$.  
Assuming a radiation-dominated cosmology with $g_\ast = 106.75$, 
the physical plasma temperature at this time is 
\begin{equation}\label{eq:TCPI}
    T_{\mathrm{phys},\CPI} = 
    \bigl( 70 \ \mathrm{TeV} \bigr) 
    \biggl( \frac{\eta_Y}{0.01 T^{-1}} \biggr) 
    \biggl( \frac{|\tilde{\mu}_{Y,5}|/T}{10^{-3}} \biggr)^2 
    \;.
\end{equation}
In other words, although the chiral asymmetry may be present in the plasma from a very early time, its effect on the hypermagnetic field does not develop until (possibly much) later when the age of the universe is comparable to 
$t_\CPI$ and the plasma has cooled to temperature $T_{\mathrm{phys},\CPI}$.  
Reducing the magnitude of the chiral asymmetry, i.e., assuming
a smaller $|\tilde{\mu}_{Y,5}|/T$ initially, delays the onset of the chiral
plasma instability.

\paragraph{\textbf{Chiral asymmetry erasure.}} 
In a relativistic electron-positron plasma described by the theory of QED, the electromagnetic charge is exactly conserved and the chiral charge is approximately conserved.  
The violation of chiral charge conservation derives from both the chiral anomaly, which leads to the phenomenon of chiral plasma instability discussed above, as well as explicit breaking induced by the nonzero electron mass.
The chiral charge changes in a scattering that converts right-chiral particles into left-chiral particles, or vice versa, and the rate for such `spin-flip' scatterings is proportional the squared electron mass $(m_e/T)^2$.  
Although the chiral charge is not exactly conserved, it is important to recognize that it is approximately conserved on time scales that are small compared to the inverse spin-flip rate.  
Similarly, the hypercharge-weighted chiral asymmetry is eventually driven to zero by scatterings involving the Yukawa couplings; the most relevant processes are Higgs decays and inverse decays with right-chiral electrons.  
The rate for these chirality-changing reactions is $\Gamma_\mathrm{f} \approx 10^{-2} y_e^2 T$, with $y_e$ electron Yukawa coupling 
$y_e = \sqrt{2} m_e / v \simeq 3 \times 10^{-6}$ 
and assuming a standard radiation-dominated cosmology, these reactions come into equilibrium when the plasma cools to a physical temperature of $T_{\mathrm{phys},\mathrm{f}} \simeq 80 \ \mathrm{TeV}$ \cite{Bodeker:2019ajh}.  
To ensure that the chiral plasma instability develops before the hypercharge-weighted chiral asymmetry is erased by Higgs decays and inverse decays, it is necessary to have $|\tilde{\mu}_{Y,5}|/T > 10^{-3}$.  
For reference, the observed baryon asymmetry of the universe today corresponds to a much smaller chemical potential of $\tilde{\mu}_{\text{\sf B}}/T \approx 10^{-8}$, but it is not unusual for large chemical potentials to be generated during the course of baryogenesis.  
New physics such as a matter-dominated phase or an injection of $e_R$ asymmetry can change the temperature of chiral asymmetry erasure; for an example, see Ref.~\cite{Chen:2019wnk}.  

\paragraph{\textbf{Magnetogenesis.}} 
As the chiral plasma instability develops, the growing helical hypermagnetic field is accompanied by a depletion of the hypercharge-weighted chiral asymmetry. 
This is because the hypercharge-weighted chiral number density $n_{Y,5} = \tilde{\mu}_{Y,5} T^2 / 6$ and the hypermagnetic helicity $\mathcal{H}_B$ are linked by the chiral anomaly, which imposes $\dot{n}_{Y,5} \propto - \alpha_Y \dot{\mathcal{H}}_{\rm M} / \pi$ \cite{Joyce:1997uy}.  
If the chiral plasma instability shuts off after the hypercharge-weighted chiral asymmetry depletes by an order one factor, the hypermagnetic helicity can be estimated as $\mathcal{H}_{{\rm M},\CPI} \sim \pi |\tilde{\mu}_{Y,5}| T_\CPI^2 / 6 \alpha_Y$.  
The coherence length and field strength are estimated as $\xi_{{\rm M},\CPI} \approx 2 \pi / (k_\CPI/2)$ and $B_\CPI \approx \sqrt{\mathcal{H}_{\rm M}/ \xi_{{\rm M},\CPI}}$, which gives $\xi_{{\rm M},\CPI} \approx (5 \times 10^5 \ \mathrm{cm}) (|\tilde{\mu}_{Y,5}|/10^{-3}T)^{-1}$ and $B_\CPI \approx (5 \times 10^{-11} \ \mathrm{G}) \, (|\tilde{\mu}_{Y,5}|/10^{-3}T)$. 
If the magnetic field evolves according to the inverse cascade scaling 
(in the fully helical case), 
$\xiM \propto t^{2/3}$ and $B \propto t^{-1/3}$ \cite{Hat84}, until recombination, 
then the physical coherence length and field strength today 
(assuming a frozen-in magnetic field and neglecting MHD dynamics at late epochs, after re-ionization)
are expected to be on the order of 
\begin{equation}
\begin{split}
    \xi_{\mathrm{M},\mathrm{phys},0} & = 
    \bigl( 9 \times 10^{-4} \ \mathrm{pc} \bigr) 
    \biggl( \frac{\eta_Y}{0.01 T^{-1}} \biggr)^{2/3} 
    \biggl( \frac{|\tilde{\mu}_{Y,5}|/T}{10^{-3}} \biggr)^{1/3}\;,
    \\ 
    B_{\mathrm{phys},0} & = 
    \bigl( 7 \times 10^{-16} \ \mathrm{G} \bigr) 
    \biggl( \frac{\eta_Y}{0.01 T^{-1}} \biggr)^{-1/3} 
    \biggl( \frac{|\tilde{\mu}_{Y,5}|/T}{10^{-3}} \biggr)^{1/3} 
    \;.
\end{split}
\end{equation}
A larger chiral asymmetry leads to a stronger magnetic field on larger
length scales today.

\paragraph{\textbf{Gravitational wave generation.}} 
The time-varying quadrupole moment of the growing hypermagnetic field
provides a source of gravitational wave radiation \cite{Deryagin:1986qq}.
As the chiral plasma instability develops, most of the magnetic 
energy is carried by the modes with coherence length
$\xi_{{\rm M},\CPI}$ (i.e., the magnetic energy is characterized by 
a spectrum that peaks at wave number $k_I \simeq 2\pi/\xi_{{\rm M},\CPI}$). 
As long as the magnetic field is still growing, however, the induced
gravitational wave spectrum peaks at the characteristic wave number
$k=2/t_\CPI=\eta_Y k_\CPI^2 / 2$ \cite{Brandenburg:2021aln}.
For $\eta_Y k_\CPI / 2 <1$, this wave number is below the cutoff wave number
for gravitational waves, $k_\CPI$.
Above this wave number, very little gravitational wave energy is produced
by the chiral plasma instability \cite{Brandenburg:2021aln}.
The gravitational wave cutoff frequency is
$f_{\rm GW} \simeq 2 k_I/(2\pi) \simeq  2/\xi_{{\rm M},\CPI}$
(the factor ``2'' is due to the quadratic nature of the source).
Since the gravitational waves' comoving frequency remains constant, the physical frequency today 
corresponds to $f_{\mathrm{GW},0} = 2 / \xi_{{\rm M},\CPI}$.
Once the CPI stops and $\tilde{\mu}_{Y,5}$ becomes depleted, the low
wave number part of the gravitational wave spectrum becomes shallower
and the peak moves toward smaller wave numbers.

The energy density carried by the gravitational waves is estimated as 
$\mathcal{E}_\mathrm{GW} \sim (G/2 \pi) a_\CPI^{-2} \xi_{{\rm M}, \CPI}^2 B_\CPI^4$.
This estimate follows from deriving the energy density $\mathcal{E}_\mathrm{GW}({\bm x},t)$ in
the standard way\footnote{In physical space, we have
${\mathcal{E}}_\mathrm{GW}^{\rm (phys)}({\bm x}_{\rm phys},t_{\rm phys})
 = \langle \partial_{t_{\rm phys}} {h}^{\rm (phys)}_{ij}({\bm x}_{\rm phys},t_{\rm phys})
 \partial_{t_{\rm phys}} {h}^{\rm (phys)}_{ij}({\bm x}_{{\rm phys}},t_{\rm phys})
 \rangle /(32 \pi G)$ where $h_{ij}^{(\rm phys)}({\bm x}_{\rm phys},t_{\rm phys)})$ is the transverse and traceless tensor
mode of the metric perturbations, and using the gravitational wave
equation $\partial_{t_{\rm phys}}^2 h_{ij}^{(\rm phys)} - \nabla^2_{\rm phys} h_{ij}^{(\rm phys)} = 
16 \pi G T_{ij}^{(\rm phys)}$ where $T_{ij}^{({\rm phys})} \sim B_i^{\rm(phys)} B_j^{\rm(phys)}$ is the transverse and traceless part of the
anisotropic part of the magnetic field stress-energy tensor, to estimate the field amplitude \cite{Gogoberidze:2007an,RoperPol:2018sap}.} 
\cite{Maggiore:1999vm}.
Next we define $\Omega_\mathrm{GW} = \mathcal{E}_\mathrm{GW} / \mathcal{E}_\mathrm{cr}$ to be the gravitational wave energy fraction today.  

Numerical estimates give
\begin{equation}\label{eq:GW_expected}
\begin{split}
    f_{\mathrm{GW},0} & = 
    \bigl( 1 \times 10^{5} \ \mathrm{Hz} \bigr) 
    \biggl( \frac{|\tilde{\mu}_{Y,5}|/T}{10^{-3}} \biggr)^{},
    \\ 
    \Omega_\mathrm{GW} h_0^2 & = 
    \bigl( 7 \times 10^{-39} \bigr) 
    \biggl( \frac{\eta_Y}{0.01 T^{-1}} \biggr)^{2} 
    \biggl( \frac{|\tilde{\mu}_{Y,5}|/T}{10^{-3}} \biggr)^{6}  
    \;.
\end{split}
\end{equation} 
A larger hypercharge-weighted chiral asymmetry moves the peak of the gravitational wave spectrum to higher frequencies (since the chiral plasma instability develops earlier) and increases the gravitational wave strength.  
For reference, the LIGO-Virgo-KAGRA gravitational wave interferometer
array is sensitive to a stochastic gravitational wave background at the
level of $\sim 10^{-7}$ for frequencies of $\sim 10$--$100 \ \mathrm{Hz}$
\cite{LIGOScientific:2016jlg}.
The future space-based detectors such as the Laser Interferometer Space Antenna (LISA) will push this sensitivity
down to $\sim 10^{-12}$ at frequencies of $\sim 1$--$10 \ \mathrm{mHz}$
\cite{Bartolo:2018qqn,Baker:2019nia,Caprini:2019pxz}. 
At still lower frequencies of $\sim 10^{-9}$--$10^{-7} \ \mathrm{Hz}$,
pulsar timing arrays (PTAs), such as Parkes PTA (PPTA) \cite{Reardon:2023gzh}, 
European PTA (EPTA)
\cite{Antoniadis:2023ott}, 
North American Nanohertz Observatory for Gravitational Waves (NANOGrav) \cite{NANOGrav:2023hvm},  Chinese PTA (CPTA) \cite{Xu:2023wog}, Indian PTA (InPTA) 
\cite{ChandraJoshi:2022etw}, and MeerKAT Pulsar Timing Array (MPTA) \cite{Miles:2022lkg} 
are sensitive to a stochastic gravitational wave
background at the level of $\Omega_\mathrm{GW} h_0^2 \sim 10^{-10}$.
Various strategies for probing higher-frequency gravitational waves,
even up to the $\mathrm{GHz}$ band, have been explored in recent years;
see Ref.~\cite{Aggarwal:2020olq} for a review of these activities.
Nevertheless, a detection of gravitational wave radiation at the level
expected here, even for $|\tilde{\mu}_{Y,5}|/T \approx 1$, seems far
out of reach.

\paragraph{\textbf{Baryon number overproduction.}} 
The presence of a helical hypermagnetic field in the early universe is expected to give rise to a baryon asymmetry \cite{Fujita:2016igl,Kamada:2016eeb,Kamada:2016cnb}.  
This is because time-varying hypermagnetic helicity sources baryon and
lepton number through the electroweak anomaly \cite{Giovannini:1997eg}.
Specifically, the conversion of a hypermagnetic field into an electromagnetic
field at the electroweak epoch at $T_\mathrm{phys} \approx 100 \ \mathrm{GeV}$
sources baryon number after the electroweak sphaleron has gone
out of equilibrium, leading to a boost in the baryon asymmetry
\cite{Kamada:2016cnb}.

The baryon number can easily be over-produced if the magnetic field
strength is too large.
Avoidance of this baryon-number overproduction imposes an upper bound
of $|\tilde{\mu}_{Y,5}|/T \lesssim 10^{-2}$ \cite{Domcke:2022uue}.
This bound is somewhat uncertain as the baryon production calculation
depends on a detailed modeling of magnetic field evolution at the
Standard Model electroweak crossover \cite{Kamada:2016cnb}, which is
not well understood.

\section{Numerical simulations}
In order to validate the preceding estimates, we have performed
three-dimensional numerical simulations using the {\sc Pencil Code}
\cite{PC}.
These simulations allow us to study the growth and evolution of the
magnetic field during the chiral plasma instability and to evaluate the
spectrum of the resulting gravitational wave radiation.

We model the Standard Model matter and radiation as a single component
plasma of charged particles interacting with the hypermagnetic field.
Several properties of the plasma are relevant to the evolution: 
the magnetic diffusivity 
(for simplicity here and below we suppress the subscript ``$Y$'')
$\eta(t) = 1 / \sigma(t)$,
the kinematic viscosity $\nu(t)$, 
the chiral diffusion coefficient $D_5(t)$, 
the chiral depletion parameter $\lambda(t)$, and 
the chiral chemical potential 
$\mu_{50} \equiv \mu_5(\xx,0) = 2 \alpha \tilde{\mu}_{5}/\pi$
that enters as an initial condition.
One can calculate $\sigma$, $\eta$, $\nu$, and $D_5$ from first principles using Standard Model particle physics.  
The hypercharge conductivity is predicted to be
$\sigma \sim T / \alpha \approx 100 \, T$ \cite{Arnold:2000dr} implying
$\eta\approx 0.01 T^{-1}$, and we assume for simplicity $\eta=\nu=D_5$.
The chiral depletion parameter $\lambda$ arises from the Standard
Model chiral anomalies, and past studies have obtained the
prediction $\lambda = 192 \, \alpha^2 / T^2 \simeq 0.02 \, T^{-2}$
\cite{Rogachevskii:2017uyc,Brandenburg:2017rcb}.
The initial chiral chemical potential can be written as 
$\mu_{50}
\approx (6 \times 10^{-6} \, T) (\tilde{\mu}_{5}/T/10^{-3})$ by fiducializing to $\tilde{\mu}_{5}/T = 10^{-3}$. 

Given the limited dynamic range of numerical simulations, it is
not possible to set the parameters, $\eta$, $\nu$, $D_5$, $\lambda$,
and $\mu_{50}$, equal to the Standard Model predictions.
Instead we consider sets of simulations with different parameters.
They can be distinguished by the relative ordering of the characteristic
quantities $v_\lambda=\mu_{50}/(\mathcal{E}_{\mathrm{cr}}\lambda)^{1/2}$ and $v_\mu=\mu_{50}\eta$.
We consider runs in regimes~I (where $v_\lambda>v_\mu$) and II (where $v_\lambda<v_\mu$).

The simulations solve a coupled system of partial differential equations
that account for MHD and the CME \cite{Brandenburg:2017rcb} to determine
the evolution of the magnetic field ${\bm B}({\bm x},t)$, the energy
density of the plasma $\rho({\bm x},t)$, the plasma velocity ${\bm u}({\bm x},t)$,
and the chiral chemical potential $\mu_5({\bm x},t)$.
In the following, we solve the following set of equations \cite{Rogachevskii:2017uyc}
\begin{eqnarray}
{\partial\AAA\over\partial t}&\!\!=\!\!&\uu\times\BB+\eta(\mu_5\BB-\JJ),
\label{dAdt}\\
{\partial\mu_5\over\partial t}&\!\!=\!\!&-\nab\cdot(\mu_5\uu)
-\lambda\eta(\mu_5\BB-\JJ)\cdot\BB+D_5\nabla^2\mu_5,\;\;
\label{dmudt}
\end{eqnarray}
\begin{eqnarray}
{\DD\uu\over\DD t}&\!\!=\!\!&{2\over\rho}\nab\cdot\left(\rho\nu\SSSS\right)-{1\over4}\nab\ln\rho
+{\uu\over3}\left(\nab\cdot\uu+\uu\cdot\nab\ln\rho\right)
\nonumber \\
&\!\!\!\!&-{\uu\over\rho}\left[\uu\cdot(\JJ\times\BB)+\eta \JJ^2\right]
+{3\over4\rho}\JJ\times\BB,
\label{dudt} \\
{\partial\ln\rho\over\partial t}
&\!\!=\!\!&-\frac{4}{3}\left(\nab\cdot\uu+\uu\cdot\nab\ln\rho\right)
\label{dlnrhodt} \nonumber \\
&\!\!\!\!&
+\frac{1}{\rho}\left[\uu\cdot(\JJ\times\BB)+\eta \JJ^2\right]
,
\end{eqnarray}
where ${\sf S}_{ij}=(\partial_j u_i+\partial_i u_j)/2-\delta_{ij}\nab\cdot\uu/3$
are the components of the rate-of-strain tensor.
We solve \Eqss{dAdt}{dlnrhodt} using the {\sc Pencil Code}
\cite{PencilCode:2020eyn}, which is a massively parallel MHD code using
sixth-order finite differences and a third-order time stepping scheme.

For discussion of the simulation results, we employ ``code units.''  
Times are measured in units of $t_\ast = 1/H_\ast$, lengths in
units of $l_\ast = c/H_\ast$, and energies in units of
$E_\ast = \mathcal{E}_{\mathrm{cr},\ast} l_\ast^3$.
Setting $\hbar=c=k_B=1$ we define $H_\ast$ by the relation
$H_\ast = \sqrt{(8\pi G/3)(\pi^2/30)(g_{\ast} T_\ast^4 / a_\ast^2)}$
where $g_{\ast} = 100$, $T_{\mathrm{phys},\ast} = T_\ast / a_\ast =
100 \, \mathrm{TeV}$, and $T_\ast = (g_{\ast S,0}/g_{\ast S,\ast})^{1/3}
T_0 \simeq 8 \times 10^{-5} \, \mathrm{eV}$.
If the CPI develops at a physical plasma temperature of
$100 \, \mathrm{TeV}$ then the age of the universe is $\sim 1/H_\ast$
and Hubble-scale Fourier modes have $k \sim H_\ast$.
The linearized gravitational wave equations are solved in wave number
space \cite{RoperPol:2018sap},
\begin{equation}
\frac{\partial^2}{\partial t^2} \tilde{h}_{+/\times} (\kk, t) 
+k^2\tilde{h}_{+/\times} (\kk, t) = {6\,H_\ast\over t \,\mathcal{E}_{\mathrm{cr}}} \tilde{T}_{+/\times}(\kk,t),
\label{GW4}
\end{equation}
where $\tilde{h}_{+/\times}=\eee_{ij}^{+/\times} (\PPP_{il}\PPP_{jm}-\half
\PPP_{ij}\PPP_{lm})\, \tilde{\hhh}_{lm}(\kk,t)$ are the Fourier-transformed
$+$ and $\times$ modes of  $\hh$, with $\eee^+_{ij} (\kk)\,\,=\,e_i^1 e_j^1 -
e_i^2 e_j^2$ and $\eee^\times_{ij} (\kk) \,=\,e_i^1 e_j^2 + e_i^2 e_j^1$
being the linear polarization basis,  $\ee^1$ and $\ee^2$ are unit
vectors perpendicular to $\kk$ and perpendicular to each other, and
$\PPP_{ij}(\kk) = \delta_{ij}-k_i k_j$ is the projection operator.
$\tilde{T}_{+/\times}$ are defined analogously.
We solve \Eq{GW4} accurate to second order in the time step and
use $1024^3$ mesh points in all of our calculations. 
Our initial conditions have a weak seed magnetic field and vanishing
plasma velocities, and the chiral chemical potential is homogeneous and
equal to the value given above.
At each time step, we calculate the spectrum of gravitational
wave radiation by solving the gravitational wave equation
sourced by the stress-energy of the plasma and magnetic field; see
Ref.~\cite{RoperPol:2018sap} for details regarding our computational
approach.

\begin{table*}\caption{
Summary of Runs discussed in this paper.
Runs~B1, B10, A1, and A12 of Ref.~\cite{Brandenburg:2021aln}
are included for comparison. 
In the last row, theoretically expected values are listed where $\eta_2 = \eta_Y / (0.01 T^{-1})$ and $\mu_3 = \tilde{\mu}_{Y,5} / 10^{-3} T$.
}\vspace{2pt}\centerline{\begin{tabular}{cccccccccccccc}
Run &$\eta H_\ast$ & 
$(\mathcal{E}_{\mathrm{cr}}\lambda)^{1/2} / H_\ast$ 
&$\mu_{50}/H_\ast$ & $\vmu$ & $\vlam$ & $\eta\mu_{50}^2/H_\ast$ & $k_1/H_\ast$ & 
$\EEM^{\max}/{\mathcal E}_{\rm cr}$ & $\EEGW^{\rm sat}/{\mathcal E}_{\rm cr}$ & $q$ \\[.7mm]
\hline\\[-3mm]
B1&$1\times10^{-6}$&$2\times10^{4}$&$10^{4}$&$1\times10^{-2}$&$5\times10^{-1}$&$1\times10^{2}$&$1\times10^{2}$& $1.6\times10^{-2}$&$4.7\times10^{-12}$&$0.027$\\
B10&$1\times10^{-3}$&$2\times10^{4}$&$10^{4}$&$1\times10^{1}$&$5\times10^{-1}$&$1\times10^{5}$&$1\times10^{2}$& $6.0\times10^{-2}$&$6.0\times10^{-9}$&$  12$\\
\hline
A1&$1\times10^{-6}$&$5\times10^{4}$&$10^{4}$&$1\times10^{-2}$&$2\times10^{-1}$&$1\times10^{2}$&$1\times10^{2}$& $4.6\times10^{-3}$&$8.9\times10^{-14}$&$0.032$\\
A12&$5\times10^{-3}$&$5\times10^{4}$&$10^{4}$&$5\times10^{1}$&$2\times10^{-1}$&$5\times10^{5}$&$5\times10^{1}$& $9.2\times10^{-3}$&$3.0\times10^{-10}$&$  18$\\
\hline
X1&$5\times10^{-8}$&$10^{10}$&$10^{6}$&$5\times10^{-2}$&$1\times10^{-4}$&$5\times10^{4}$&$5\times10^{3}$&  $2.4\times10^{-9}$&$8.8\times10^{-31}$&$0.39$\\
X2&$5\times10^{-9}$&$10^{10}$&$10^{6}$&$5\times10^{-3}$&$1\times10^{-4}$&$5\times10^{3}$&$5\times10^{3}$&  $2.4\times10^{-9}$&$1.6\times10^{-30}$&$0.53$\\
X3&$5\times10^{-10}$&$10^{10}$&$10^{6}$&$5\times10^{-4}$&$1\times10^{-4}$&$5\times10^{2}$&$5\times10^{3}$& $2.4\times10^{-9}$&$1.1\times10^{-30}$&$0.44$\\
X4&$5\times10^{-11}$&$10^{10}$&$10^{6}$&$5\times10^{-5}$&$1\times10^{-4}$&$5\times10^{1}$&$5\times10^{3}$& $2.3\times10^{-9}$&$3.1\times10^{-31}$&$0.12$\\
\hline
Y1&$5\times10^{-8}$&$7\times10^{11}$&$10^{6}$&$5\times10^{-2}$&$1\times10^{-6}$&$5\times10^{4}$&$5\times10^{3}$& $4.9\times10^{-13}$&$3.6\times10^{-38}$&$0.39$\\
Y2&$5\times10^{-8}$&$7\times10^{11}$&$10^{6}$&$5\times10^{-2}$&$1\times10^{-6}$&$5\times10^{4}$&$2\times10^{3}$& $4.4\times10^{-13}$&$3.2\times10^{-37}$&$1.3$\\
Y3&$5\times10^{-8}$&$7\times10^{11}$&$10^{6}$&$5\times10^{-2}$&$1\times10^{-6}$&$5\times10^{4}$&$1\times10^{3}$& $3.3\times10^{-13}$&$6.9\times10^{-37}$&$2.5$\\ 
\hline
expected &$10^{-15}\eta_2$&$6\times10^{12}$&$5\times10^{7}\mu_3$&$6\times10^{-8}\eta_2\mu_3$&$8\times10^{-6}\mu_3$&$3\eta_2\mu_3^2$&---&$6\times10^{-15}\mu_3^2$&$7\times10^{-39}\eta_2^2\mu_3^6$& --- \\
\hline
\label{Tsummary}\end{tabular}}\end{table*}

In \Tab{Tsummary}, we summarize the parameters used in our simulations,
including the smallest wavenumber $k_1$, and the key results.
We consider two series of runs that we refer to as X and Y.
We also compare with two pairs of runs, A1 and A12, as well as B1 and B10,
both from Ref.~\cite{Brandenburg:2021aln}.
where $\mu/v_\lambda$ increases from 0.02 and 0.05 to 20 and 250, respectively.
Also the efficiency parameters increases from 0.03 to 12 and 18, respectively.
The runs of series~X and Y are subdivided further into Runs~X1--X4 and Runs~Y1--Y3.
Our runs of series~X have increasing values of $v_\mu$ and cross from
regime~II (for Run~X1) into regime~I (for Run~X4).
For the runs of series~X, we take $\eta = \nu = D_5 = 5\times10^{-11}/H_\ast$,
$\mathcal{E}_{\mathrm{cr}}\lambda=10^{20}\,H_\ast^2$,
and $\mu_{50}=10^6\,H_\ast$.
We also give the efficiency of gravitational wave production,
\EQ
q=(k_{\rm peak}/H_\ast)\left.\sqrt{\EEGW^{\rm sat}{\mathcal E}_{\rm cr}}\right/\EEM^{\max},
\label{qdef2}
\EN
where we estimate $k_{\rm peak}=k_\mu\min(1,\vmu/\vlam)$ 
where $k_\mu = \mu_{50}/2$.  
This means that $k_{\rm peak}=k_\mu$ when $\vmu > \vlam$ (regime~II)
and $k_{\rm peak}=k_\lambda/4$ when $\vmu < \vlam$ (regime~I); see
Ref.~\cite{Brandenburg:2017rcb}.

In the last row of \Tab{Tsummary}, we compare with the theoretically expected values.
Obviously, our values of $\eta$ are about seven orders of magnitude too large.
This reflects the fact that our simulations are unable to capture a sufficiently large range of length scales.
Consequently, also our values of $v_\mu$ are by about six orders of magnitude too large.
Furthermore, $\eta\mu_{50}^2/H_\ast$ is by about four orders of magnitude too large.
Most of the other simulation values in the table are not so far from the theoretically expected values.

\begin{figure}[t]
\centering
\includegraphics[width=\columnwidth]{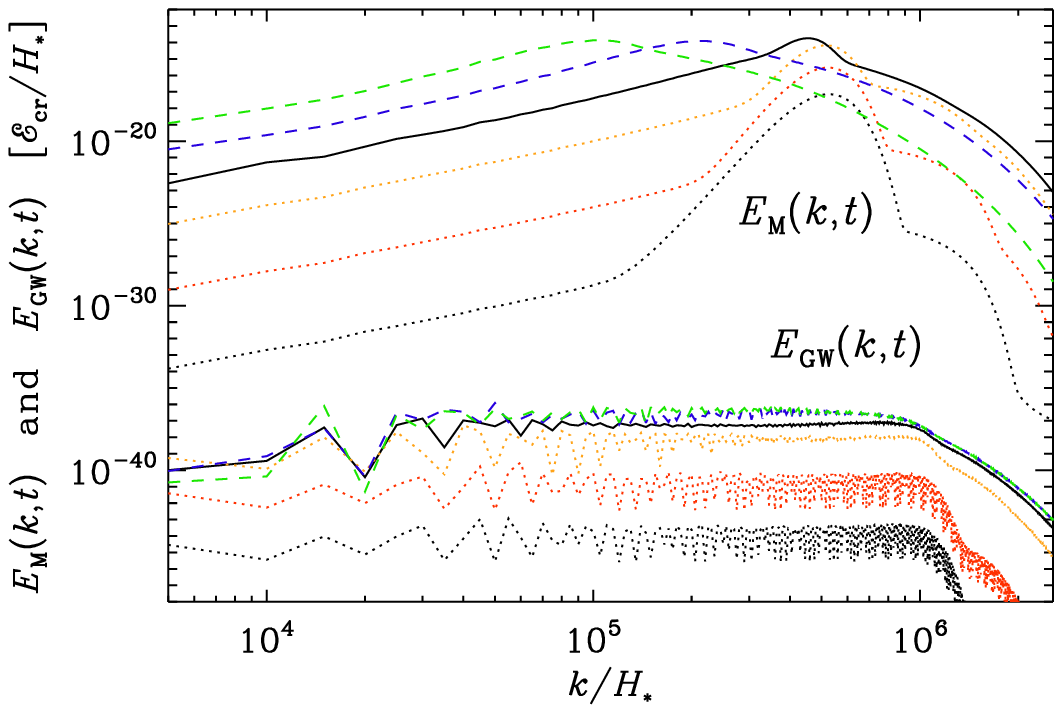}
\caption{\label{fig:spectra}
Spectra (per linear wave number interval) of magnetic energy
$\EM(k,t)$ (upper curves) and gravitational wave energy $\EGW(k,t)$
(lower curves) from the chiral plasma instability and turbulent MHD
evolution for Run~X4, where $\mu_{5}=10^{6}\,H_\ast$, $\mathcal{E}_{\mathrm{cr}}\lambda=10^{20}/H_\ast$,
$\eta=5\times10^{-11}\,H_\ast$ which implies $v_\lambda=10^{-4}$
and $v_\mu=5\times10^{-5}$ (corresponding to regime~I).
The solid curves are for $tH_\ast=2.98$, when $\EEM$ is maximum.
The dotted curves are for $tH_\ast=2.41$ (black), $2.56$ (red), and $2.71$
(orange), before $\EEM$ is maximum, while the dashed curves are for $tH_\ast=3.66$ (blue)
and $tH_\ast=5.37$ (green), when $\EEM$ is decaying.
}\end{figure}

The evolution of the magnetic and gravitational wave energy spectra for
Run~X4 is shown in Fig.~\ref{fig:spectra}.
The energy densities may be written as
$\mathcal{E} = \int_0^\infty \! \mathrm{d}k \, E(k)$ where
$k$ is the wave number and $E(k)$ is the energy spectrum.
For the parameters of Run~X4, the instability length scale corresponds
to a wavenumber of $k_\CPI = 10^6 H_\ast$, which agrees with the wave
number above which $\EGW(k)$ drops sharply.
The instability time scale normalized to the Hubble time is $t_\CPI H_\ast = 0.08$,
which is about 100 times longer than the time step.
The magnetic energy spectrum grows initially for modes with $k \approx k_\CPI/2 = 5 \times 10^5 H_\ast$
(see the upper set of dotted lines in Fig.~\ref{fig:spectra}).
Later, the peak evolves to smaller $k$ with an inverse cascade scaling,
which is consistent with earlier simulations \cite{Brandenburg:2017rcb}.
The generated magnetic field is then maximally helical; see Fig.~8(b)
of Ref.~\cite{Brandenburg:2021bfx}.

The gravitational wave energy spectra grow in time as long as the
magnetic energy has not yet reached its maximum.
In this phase, as discussed above, the gravitational wave
spectrum is expected to peak at the characteristic wave number
$k = 2 / t_\CPI = \eta k_\CPI^2 / 2 = 25\,H_\ast$, which is here much smaller than
$k_\CPI = 10^6 H_\ast$, 
but larger than the horizon wave number, $k = H_\ast$.
When the magnetic energy density has reached its maximum value, the
gravitational wave spectrum has nearly saturated and is then approximately
independent of $k$ for $k < k_\CPI/2$.
In principle, it is possible to have a declining $k^{-2}$ spectrum in
the range $\eta k_\CPI^2 / 2 \leq k \leq k_\CPI$, but this is only seen
in our models with larger diffusivity.
The absence of a $k^{-2}$ subrange in the gravitational wave spectrum
could also be an artifact of insufficient numerical resolution.
In any case, once the gravitational wave spectrum saturates,
we would expect the development of a flat ($\EGW \propto k^0$) spectrum.
Such a flat spectrum is expected to extend all the way to the horizon
wave number $k = H_\CPI$ \cite{RoperPol:2019wvy, RoperPol:2022iel, Sharma:2022ysf}.
Therefore, the total gravitational wave energy is expected to be
proportional to $k_\CPI/2-H_\ast$.
However, since $k_\CPI/2$ is already much larger than our lower cutoff
value $k_1$, the error in our estimate of $\EEGW\propto k_\CPI/2-k_1$
is negligible.

\begin{figure}[t]
\centering
\includegraphics[width=\columnwidth]{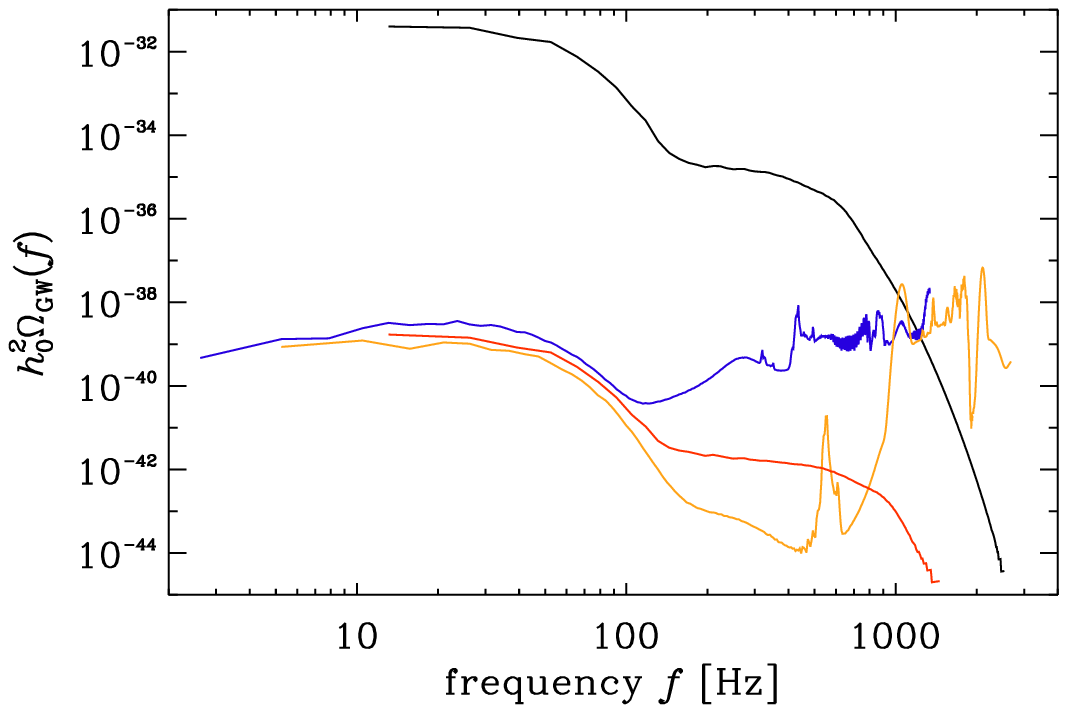}
\caption{\label{fig:pspecm_D1024_1e5_1e6_49e22}
Comparison of $h_0^2\Omega_{\rm GW}(f)$ versus $f$ for runs with
$k_1/H_\ast = 10^{3}$ (Run~Y3, blue), $2 \times 10^{3}$ (Run~Y2, orange),
and $5 \times 10^{3}$ (Run~Y1, red), with
$\lambda = 49 \times 10^{22} H_\ast^2 / \mathcal{E}_{\mathrm{cr},\ast}$
and $\eta = 5 \times 10^{-8} H_\ast^{-1}$
with $H_0 = 100 \, h_0 \km \s^{-1} \Mpc^{-1}$.
Run~X4 with with $\lambda = 10^{20} H_\ast^2 / \mathcal{E}_{\mathrm{cr},\ast}$
and again $\eta = 5 \times 10^{-8} H_\ast^{-1}$ is
shown as the black line for reference.
}\end{figure}

In \Fig{fig:pspecm_D1024_1e5_1e6_49e22}, we show gravitational wave
spectra for a few runs with smaller values of the minimum wave number
in the simulations.
We see that the spectra remain nearly flat, but the spectra are also
becoming more irregular at large wave numbers.
This is likely an artifact of insufficient numerical resolution.
We also see that most of the gravitational wave energy is at
frequencies below about $1\,\kHz$, but this value would increase
with increasing values of $\mu_{50}$, beyond the value of $10^6 H_\ast$
adopted here.
The fiducial value of $\tilde{\mu}_{Y,5} = 10^{-3} T$
in \Eq{eq:GW_expected} corresponds to $\mu_{50} \approx 5
\times 10^7 \, H_\ast$, and since the simulations presented in
\Fig{fig:pspecm_D1024_1e5_1e6_49e22} have $\mu_{50} = 10^6 H_\ast$,
the gravitational wave frequencies are proportionally smaller.

\begin{figure}[t!]\begin{center}
    \includegraphics[width=\columnwidth]{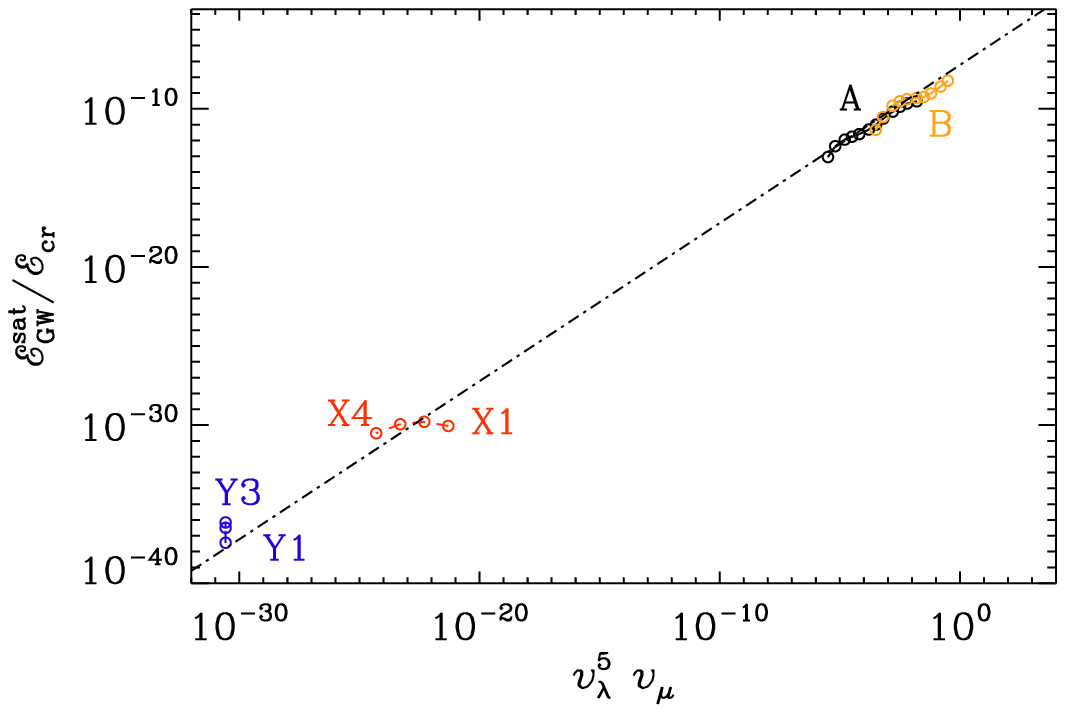}
\end{center}\caption[]{
Dependence of $\EEGW^{\rm sat}$ on $\vlam^5\vmu$ for our runs of Series~X (red) and Y (blue),
as well as Series~A (black) and B (orange) of Ref.~\cite{Brandenburg:2021aln}.
}\label{fig:pall}
\end{figure}

Earlier work showed that $\EEGW^{\rm sat}$ grows approximately linearly
with $\eta$ and was proportional to $(\mathcal{E}_{\mathrm{cr}}\lambda)^{-5/2}$, which
leads to the combined dependence \cite{Brandenburg:2021aln}
\EQ
\EEGW^{\rm sat}/{\mathcal E}_{\rm cr}\approx6\times10^{-8}\,\vlam^5\vmu,
\label{vlam5vmu}
\EN
which implies that $\EEGW^{\rm sat}\propto\mu_{50}^6$.
In \Fig{fig:pall}, we plot $\EEGW^{\rm sat}$ versus $\vlam^5\vmu$
for Runs~X1--X4 and Y1--Y3.
We see that \Eq{vlam5vmu} agrees reasonably well with
our numerical data.
Compared with the runs of Ref.~\cite{Brandenburg:2021aln}, the new
one in Fig.~\ref{fig:spectra} has much smaller values of $v_\lambda$
(here $v_\lambda=10^{-4}$ instead of 0.5 for the old runs of Series~B)
and $v_\mu$ (here $v_\mu=5\times10^{-5}$ instead of $10^{-2}$, which
was their smallest value).
This has been achieved by having $k_\CPI$ much larger (here $10^6$
instead of $10^4$, for example).
This also means that we have to choose a correspondingly larger
value of the minimum wave number, $k_1$.

\begin{figure}[t]
    \centering
    \includegraphics[width=\columnwidth]{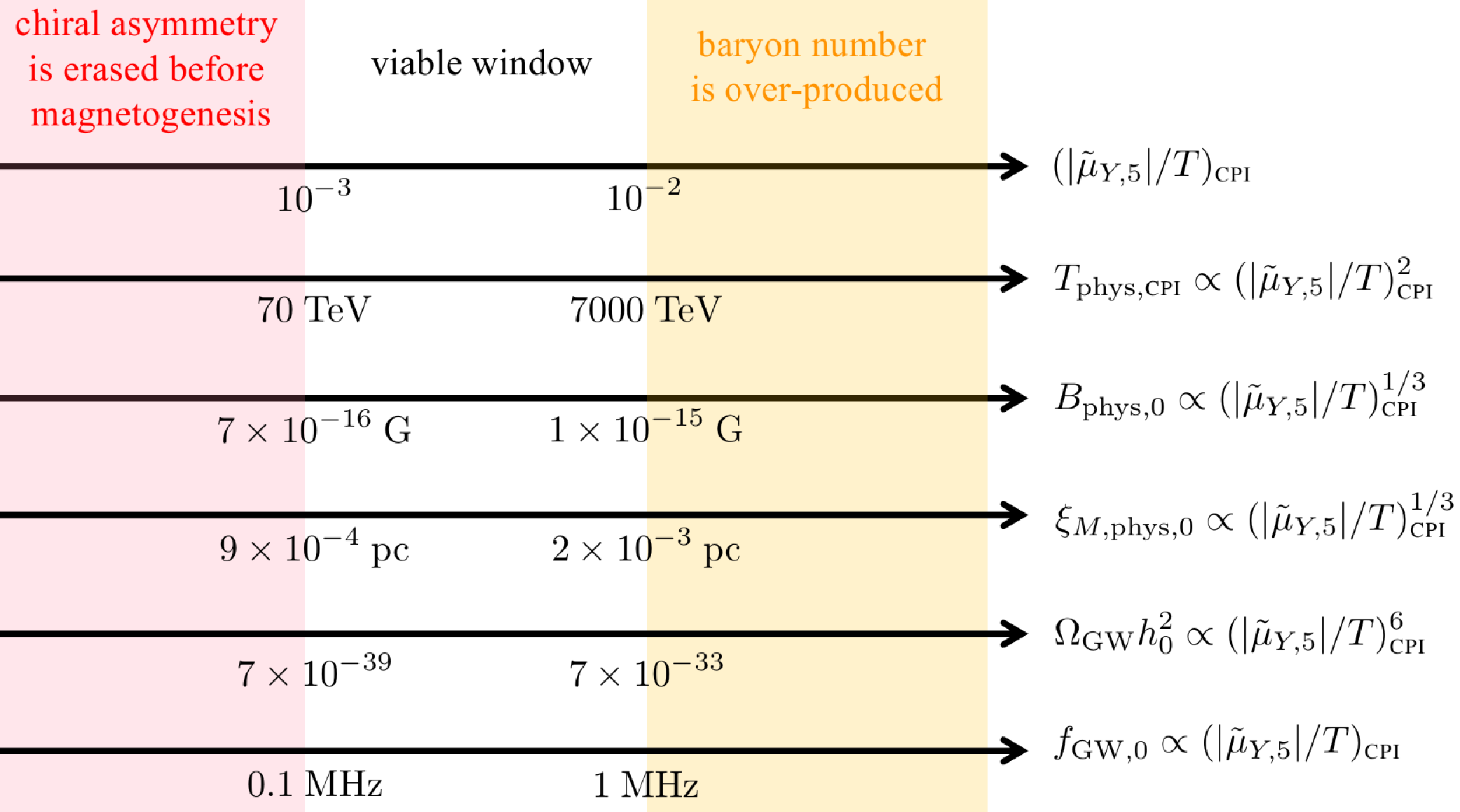}
    \caption{\label{fig:summary}
    Summary of the viable parameter space for the hypercharge-weighted chiral chemical potential $\tilde{\mu}_{Y,5}$ and predictions for the relic magnetic field and gravitational wave radiation.  
    }
\end{figure}

\section{Conclusions}
Our estimates of the key variables are summarized in Fig.~\ref{fig:summary}. 
Since we assume Standard Model particles and interactions, as well as a standard cosmology with radiation domination at temperatures $T > 100 \ \mathrm{TeV}$, the observables depend only on the single dimensionless parameter $|\tilde{\mu}_{Y,5}|/T$, which controls the size of the 
initial hypercharge-weighted chiral asymmetry.  
To ensure that the instability develops before the chiral asymmetry is washed out by reactions such as Higgs decays and inverse decays, we need $|\tilde{\mu}_{Y,5}|/T \gtrsim 10^{-3}$.  
On the other hand, to avoid over-producing the baryon asymmetry we need $|\tilde{\mu}_{Y,5}|/T \lesssim 10^{-2}$.  
This leaves an approximately one-decade wide window of viable parameter space.  
The predicted magnetic field strength today, assuming inverse cascade scaling from production until recombination,
is at the level of $10^{-15} \ \mathrm{Gauss}$.  
An intergalactic magnetic field at this level is strong
enough to explain observations of distant TeV blazars, which provide
evidence for a nonzero intergalactic magnetic field at the level
$\gtrsim 10^{-16} \ \mathrm{Gauss}$ \cite{Neronov:2010gir},
although the predicted coherence length is too small.
Alternative models with CPI-like equations of motion and late-time
dynamics \cite{Zhitnitsky:2019ijg,Basilakos:2019acj} may lead to stronger
large-scale fields.
The same magnetic field may help to explain the origin of galactic magnetic fields by providing a seed for the galactic dynamo.  
The strength of the gravitational wave signal is expected to depend strongly on the value of $|\tilde{\mu}_{Y,5}|/T$, going as its sixth power.
The typical frequency of this signal is expected to fall near
$\sim \mathrm{GHz}$, putting it into a frequency band that is being targeted
by several recently proposed probes of high-frequency gravitational
wave radiation.
However, within the viable window, the gravitational wave signal is likely far too weak for detection.  

{\bf Data availability}---The source code used for the
simulations of this study, the {\sc Pencil Code},
is freely available from Ref.~\cite{PC}.
The simulation setups and the corresponding data
are freely available from Ref.~\cite{DATA}.

\vspace{2mm}
{\bf Acknowledgements}---We thank Kohei Kamada, Sayan Mandal, and Jonathan Stepp for useful discussions and comments. 
A.J.L.\ is grateful to the Mainz Institute for Theoretical Physics (MITP)
of the DFG Cluster of Excellence PRISMA+ (Project ID 39083149)
as well as to the Aspen Center for Physics, which is supported by National
Science Foundation grant PHY-2210452, for their hospitality during the
completion of this work.
Support through the NASA ATP award 80NSSC22K0825, the Swedish  Research
Council, grant 2019-04234, and Shota Rustaveli GNSF (grant FR/18-1462)
are gratefully acknowledged.
We acknowledge the allocation of computing resources provided by the
Swedish National Allocations Committee at the Center for Parallel
Computers at the Royal Institute of Technology in Stockholm.
E.C.\ acknowledges the Pake Fellowship, and G.S.\ acknowledges
support from the Undergraduate Research Office in the form of a Summer
Undergraduate Research Fellowships (SURF) at Carnegie Mellon University.

\bibliographystyle{h-physrev5}
\bibliography{Chirality-BBN}

\begin{thebibliography}{10}

\bibitem{Kolb:1979qa}
E.~W. Kolb and S.~Wolfram,
\newblock Nucl. Phys. B {\bf 172}, 224 (1980),
\newblock [Erratum: Nucl.Phys.B 195, 542 (1982)].

\bibitem{Fukugita:1986hr}
M.~Fukugita and T.~Yanagida,
\newblock Phys. Lett. {\bf B174}, 45 (1986).

\bibitem{Servant:2013uwa}
G.~Servant and S.~Tulin,
\newblock (2013), arXiv:1304.3464.

\bibitem{Dick:1999je}
K.~Dick, M.~Lindner, M.~Ratz, and D.~Wright,
\newblock Phys.Rev.Lett. {\bf 84}, 4039 (2000), arXiv:hep-ph/9907562.

\bibitem{Murayama:2002je}
H.~Murayama and A.~Pierce,
\newblock Phys. Rev. Lett. {\bf 89}, 271601 (2002), arXiv:hep-ph/0206177.

\bibitem{Campbell:1990fa}
B.~A. Campbell, S.~Davidson, J.~R. Ellis, and K.~A. Olive,
\newblock Phys.Lett. {\bf B256}, 484 (1991).

\bibitem{Joyce:1997uy}
M.~Joyce and M.~E. Shaposhnikov,
\newblock Phys.Rev.Lett. {\bf 79}, 1193 (1997), arXiv:astro-ph/9703005.

\bibitem{Akamatsu:2013pjd}
Y.~Akamatsu and N.~Yamamoto,
\newblock Phys. Rev. Lett. {\bf 111}, 052002 (2013), arXiv:1302.2125.

\bibitem{Neronov:2010gir}
A.~Neronov and I.~Vovk,
\newblock Science {\bf 328}, 73 (2010), arXiv:1006.3504.

\bibitem{Vachaspati:2020blt}
T.~Vachaspati,
\newblock Rept. Prog. Phys. {\bf 84}, 074901 (2021), arXiv:2010.10525.

\bibitem{Deryagin:1986qq}
D.~V. Deryagin, D.~Y. Grigoriev, V.~A. Rubakov, and M.~V. Sazhin,
\newblock Mod. Phys. Lett. A {\bf 1}, 593 (1986).

\bibitem{Brandenburg:2021aln}
A.~Brandenburg, Y.~He, T.~Kahniashvili, M.~Rheinhardt, and J.~Schober,
\newblock Astrophys. J. {\bf 911}, 110 (2021), arXiv:2101.08178.

\bibitem{Anber:2009ua}
M.~M. Anber and L.~Sorbo,
\newblock Phys. Rev. D {\bf 81}, 043534 (2010), arXiv:0908.4089.

\bibitem{Barnaby:2012xt}
N.~Barnaby {\em et~al.},
\newblock Phys. Rev. D {\bf 86}, 103508 (2012), arXiv:1206.6117.

\bibitem{Domcke:2016bkh}
V.~Domcke, M.~Pieroni, and P.~Bin\'etruy,
\newblock JCAP {\bf 06}, 031 (2016), arXiv:1603.01287.

\bibitem{Machado:2018nqk}
C.~S. Machado, W.~Ratzinger, P.~Schwaller, and B.~A. Stefanek,
\newblock JHEP {\bf 01}, 053 (2019), arXiv:1811.01950.

\bibitem{Brandenburg:2023rul}
A.~Brandenburg, K.~Kamada, and J.~Schober,
\newblock Phys. Rev. Res. {\bf 5}, L022028 (2023), arXiv:2302.00512.

\bibitem{Brandenburg:2023aco}
A.~Brandenburg, K.~Kamada, K.~Mukaida, K.~Schmitz, and J.~Schober,
\newblock Phys. Rev. D {\bf 108}, 063529 (2023), 2304.06612.

\bibitem{Bell:1969ts}
J.~S. Bell and R.~Jackiw,
\newblock Nuovo Cim. A {\bf 60}, 47 (1969).

\bibitem{Adler:1969gk}
S.~L. Adler,
\newblock Phys. Rev. {\bf 177}, 2426 (1969).

\bibitem{Vilenkin:1980fu}
A.~Vilenkin,
\newblock Phys. Rev. D {\bf 22}, 3080 (1980).

\bibitem{Boyarsky:2011uy}
A.~Boyarsky, J.~Fr{\"o}hlich, and O.~Ruchayskiy,
\newblock Phys. Rev. Lett. {\bf 108}, 031301 (2012), arXiv:1109.3350.

\bibitem{Boyarsky:2012ex}
A.~Boyarsky, O.~Ruchayskiy, and M.~Shaposhnikov,
\newblock Phys. Rev. Lett. {\bf 109}, 111602 (2012), arXiv:1204.3604.

\bibitem{Boyarsky:2015faa}
A.~Boyarsky, J.~Fr{\"o}hlich, and O.~Ruchayskiy,
\newblock Phys. Rev. D {\bf 92}, 043004 (2015), arXiv:1504.04854.

\bibitem{Brandenburg:2017rcb}
A.~Brandenburg {\em et~al.},
\newblock Astrophys. J. Lett. {\bf 845}, L21 (2017), arXiv:1707.03385.

\bibitem{Kamada:2022nyt}
K.~Kamada, N.~Yamamoto, and D.-L. Yang,
\newblock Prog. Part. Nucl. Phys. {\bf 129}, 104016 (2023), arXiv:2207.09184.

\bibitem{Kamada:2016eeb}
K.~Kamada and A.~J. Long,
\newblock Phys. Rev. D {\bf 94}, 063501 (2016), arXiv:1606.08891.

\bibitem{Kamada:2016cnb}
K.~Kamada and A.~J. Long,
\newblock Phys. Rev. D {\bf 94}, 123509 (2016), arXiv:1610.03074.

\bibitem{Bodeker:2019ajh}
D.~B\"odeker and D.~Schr\"oder,
\newblock JCAP {\bf 05}, 010 (2019), arXiv:1902.07220.

\bibitem{Chen:2019wnk}
M.-C. Chen, S.~Ipek, and M.~Ratz,
\newblock Phys. Rev. D {\bf 100}, 035011 (2019), arXiv:1903.06211.

\bibitem{Hat84}
T.~{Hatori},
\newblock JPSJ {\bf 53}, 2539 (1984).

\bibitem{Gogoberidze:2007an}
G.~Gogoberidze, T.~Kahniashvili, and A.~Kosowsky,
\newblock Phys. Rev. D {\bf 76}, 083002 (2007), arXiv:0705.1733.

\bibitem{RoperPol:2018sap}
A.~Roper~Pol, A.~Brandenburg, T.~Kahniashvili, A.~Kosowsky, and S.~Mandal,
\newblock Geophys. Astrophys. Fluid Dynamics {\bf 114}, 130 (2020),
  arXiv:1807.05479.

\bibitem{Maggiore:1999vm}
M.~Maggiore,
\newblock Phys. Rept. {\bf 331}, 283 (2000), arXiv:gr-qc/9909001.

\bibitem{LIGOScientific:2016jlg}
LIGO Scientific, Virgo, B.~P. Abbott {\em et~al.},
\newblock Phys. Rev. Lett. {\bf 118}, 121101 (2017), arXiv:1612.02029,
\newblock [Erratum: Phys.Rev.Lett. 119, 029901 (2017)].

\bibitem{Bartolo:2018qqn}
N.~Bartolo {\em et~al.},
\newblock JCAP {\bf 11}, 034 (2018), arXiv:1806.02819.

\bibitem{Baker:2019nia}
J.~Baker {\em et~al.},
\newblock (2019), arXiv:1907.06482.

\bibitem{Caprini:2019pxz}
C.~Caprini {\em et~al.},
\newblock JCAP {\bf 11}, 017 (2019), arXiv:1906.09244.

\bibitem{Reardon:2023gzh}
D.~J. Reardon {\em et~al.},
\newblock Astrophys. J. Lett. {\bf 951} (2023), arXiv:2306.16215.

\bibitem{Antoniadis:2023ott}
EPTA, InPTA:, J.~Antoniadis {\em et~al.},
\newblock Astron. Astrophys. {\bf 678}, A50 (2023), arXiv:2306.16214.

\bibitem{NANOGrav:2023hvm}
NANOGrav, A.~Afzal {\em et~al.},
\newblock Astrophys. J. Lett. {\bf 951} (2023), arXiv:2306.16219.

\bibitem{Xu:2023wog}
H.~Xu {\em et~al.},
\newblock Res. Astron. Astrophys. {\bf 23}, 075024 (2023), arXiv:2306.16216.

\bibitem{ChandraJoshi:2022etw}
B.~Chandra~Joshi {\em et~al.},
\newblock J. Astrophys. Astron. {\bf 43}, 98 (2022), arXiv:2207.06461.

\bibitem{Miles:2022lkg}
M.~T. Miles {\em et~al.},
\newblock Mon. Not. Roy. Astron. Soc. {\bf 519}, 3976 (2023), arXiv:2212.04648.

\bibitem{Aggarwal:2020olq}
N.~Aggarwal {\em et~al.},
\newblock Living Rev. Rel. {\bf 24}, 4 (2021), arXiv:2011.12414.

\bibitem{Fujita:2016igl}
T.~Fujita and K.~Kamada,
\newblock Phys. Rev. D {\bf 93}, 083520 (2016), arXiv:1602.02109.

\bibitem{Giovannini:1997eg}
M.~Giovannini and M.~E. Shaposhnikov,
\newblock Phys. Rev. D {\bf 57}, 2186 (1998), arXiv:hep-ph/9710234.

\bibitem{Domcke:2022uue}
V.~Domcke, K.~Kamada, K.~Mukaida, K.~Schmitz, and M.~Yamada,
\newblock (2022), arXiv:2208.03237.

\bibitem{PC}
The pencil code. doi:10.5281/zenodo.2315093.
  \href{https://github.com/pencil-code}{https://github.com/pencil-code}.

\bibitem{Arnold:2000dr}
P.~B. Arnold, G.~D. Moore, and L.~G. Yaffe,
\newblock JHEP {\bf 11}, 001 (2000), arXiv:hep-ph/0010177.

\bibitem{Rogachevskii:2017uyc}
I.~Rogachevskii {\em et~al.},
\newblock Astrophys. J. {\bf 846}, 153 (2017), arXiv:1705.00378.

\bibitem{PencilCode:2020eyn}
Pencil Code Collaboration, A.~Brandenburg {\em et~al.},
\newblock J. Open Source Softw. {\bf 6}, 2807 (2021), arXiv:2009.08231.

\bibitem{Brandenburg:2021bfx}
A.~Brandenburg, Y.~He, and R.~Sharma,
\newblock Astrophys. J. {\bf 922}, 192 (2021), arXiv:2107.12333.

\bibitem{RoperPol:2019wvy}
A.~Roper~Pol, S.~Mandal, A.~Brandenburg, T.~Kahniashvili, and A.~Kosowsky,
\newblock Phys. Rev. D {\bf 102}, 083512 (2020), arXiv:1903.08585.

\bibitem{RoperPol:2022iel}
A.~Roper~Pol, C.~Caprini, A.~Neronov, and D.~Semikoz,
\newblock Phys. Rev. D {\bf 105}, 123502 (2022), arXiv:2201.05630.

\bibitem{Sharma:2022ysf}
R.~Sharma and A.~Brandenburg,
\newblock Phys. Rev. D {\bf 106}, 103536 (2022), arXiv:2206.00055.

\bibitem{Zhitnitsky:2019ijg}
A.~R. Zhitnitsky,
\newblock Phys. Rev. D {\bf 99}, 103518 (2019), arXiv:1902.07737.

\bibitem{Basilakos:2019acj}
S.~Basilakos, N.~E. Mavromatos, and J.~Sol\`a~Peracaula,
\newblock Phys. Rev. D {\bf 101}, 045001 (2020), arXiv:1907.04890.

\bibitem{DATA}
A.~{Brandenburg}, E.~{Clarke}, T.~{Kahniashvili}, A.~J. {Strong}, and G.~{Sun},
\newblock {Datasets for Relic gravitational waves from the chiral plasma
  instability in the standard cosmological model, doi:10.5281/zenodo.8157463
  (v2023.07.17)}; see also
  \url{http://norlx65.nordita.org/~brandenb/proj/GWfromSM/} for easier access .

\end{thebibliography}

\end{document}